\documentclass[a4paper,twocolumn,
english,aps,prb,floatfix,showpacs]{revtex4}
\usepackage[T1]{fontenc}
\usepackage[latin1]{inputenc}
\usepackage{amsmath}
\usepackage{babel}
\usepackage{graphics}
\usepackage{amssymb}


\begin{document}
\title
{Anomalous dynamics in two-- and three-- dimensional Heisenberg-Mattis spin
glasses}
\author {S.L.A. \surname{de Queiroz}}

\email{sldq@if.ufrj.br}

\affiliation{Instituto de F\'\i sica, Universidade Federal do
Rio de Janeiro, Caixa Postal 68528, 21941-972
Rio de Janeiro RJ, Brazil}

\author {R. B. \surname{Stinchcombe}}

\email{stinch@thphys.ox.ac.uk}

\affiliation{Rudolf Peierls Centre for Theoretical Physics, University of
Oxford, 1 Keble Road, Oxford OX1 3NP, United Kingdom}

\date{\today}

\begin{abstract}
We investigate the spectral and localization properties 
of unmagnetized Heisenberg-Mattis spin glasses, in space dimensionalities
$d=2$ and $3$, at $T=0$.
We use  numerical transfer-matrix methods combined with finite-size
scaling to calculate
Lyapunov  exponents, and eigenvalue-counting theorems, coupled with
Gaussian elimination algorithms, to evaluate densities of states.
In $d=2$ we find that all states are localized, with the localization
length diverging as $\omega^{-1}$, as energy $\omega \to 0$. Logarithmic
corrections to  density of states behave in accordance with theoretical
predictions. In $d=3$ the density-of-states dependence on energy is the
same as for spin waves in pure antiferromagnets, again in agreement with
theoretical predictions, though the corresponding amplitudes differ.
\end{abstract}
\maketitle
 
\section{Introduction} 
\label{intro}
The study of low-lying magnetic excitations in quenched disordered systems
presents a number of challenges. While the absence of
translational invariance is a complicator arising in all aspects both
of static and dynamic behavior of inhomogeneous magnets, 
investigation of spin waves is made even harder because,
in many cases of interest, the exact ground state configuration is not
known.

One way around the latter obstacle has been to resort to simplified model
systems for which the exact ground state is known, but which nevertheless
still display non-trivial dynamical features. Such features, it is
expected, may shed light on the behavior of their experimentally-realized,
rather more complex, counterparts.

Here we deal with vector spin glasses, i. e., Heisenberg spins with
competing ferro-- and antiferromagnetic interactions. 
It is known that the simplest realization of the Edwards-Anderson
picture, where one has equal
concentrations of positive and negative nearest-neighbor bonds of equal
strength, leads (in lattices of space dimensionality $d>1$) to frustration
and, consequently, to a macroscopically degenerate (classical) ground
state.

The drawback just described does not arise in Mattis spin-glasses, where
the Mattis transformation~\cite{dcm76} ``gauges away'' disorder effects,
as far as most static aspects are concerned. It is known
that the Mattis transformation does not remove the disorder
effects in the dynamics of these so-called Heisenberg-Mattis spin glasses,
which is non-trivial. Indeed, investigations of spin-wave propagation
in such systems~\cite{ds77,clh77,ds79,ch79,sp88,ga04} have unveiled many
features  
which stand in stark contrast, e.g., to the Halperin-Saslow (hydrodynamic)
picture~\cite{hs77} of a linear dispersion relation for low-energy
excitations. 

Here, we shall assume that the spin magnitude is $|{\mathbf S}|
\gg 1$, so that quantum fluctuations can be safely
neglected~\cite{sp88,ga04} (classical limit). 

An alternative to using the Mattis picture can be pursued by 
studying usual spin glasses (i.e. with random $\pm J$ bonds) in the
high-field limit, as this additional feature stabilizes a 
ferromagnetic-like ground state while still incorporating quenched 
(bond) disorder~\cite{ahc92,ah93,ahc93}. However, results thus obtained 
differ rather drastically from those pertaining to the zero-field case.
In fact, it has been found that, even in zero field and space
dimensionality 
$d=1$ where frustration effects are absent, ``unmagnetized'' spin glasses 
(i.e. in which the concentrations of ferro- ($p$) and antiferromagnetic
($1-p$) bonds are equal) differ substantially from their
``magnetized'' ($p \neq 1/2$) counterparts~\cite{ew92}. 

In this paper, we investigate the spectral and localization properties of
Heisenberg-Mattis spin glasses. Our emphasis is on unmagnetized systems in
space dimensionalities $d=2$ and $3$, at $T=0$.  We use numerical
transfer-matrix methods to calculate Lyapunov
exponents~\cite{ps81,bh89,ew92}, and eigenvalue-counting theorems, coupled
with Gaussian elimination algorithms~\cite{sb83,sne86}, to evaluate
densities of states. Though early numerical studies~\cite{clh77,ch79}
already highlighted a number of distinctive features exhibited by such
systems, motivation for further research is to be found in recent
theoretical insights~\cite{gc03,ga04}, especially in
connection with the low-energy, long-wavelength regime. 
  
In Section~\ref{sec:2} we recall pertinent aspects of Heisenberg-Mattis
spin glasses. Section~\ref{sec:3} reports on an extension, to $d=2$ and
$3$, of the analytical scaling techniques introduced in
Ref.~\onlinecite{sp88} for $d=1$; in Section~\ref{sec:4} we report
numerical calculations of Lyapunov exponents and of densities of states,
for $d=2$ and $3$.  Finally, in section~\ref{sec:conc}, concluding remarks
are made. 

\section
{Heisenberg-Mattis spin glasses}
\label{sec:2}

We consider Heisenberg spins on sites of a square, or simple-cubic, 
lattice, with nearest-neighbor couplings:
\begin{equation}
{\cal H}= -\sum_{\langle i,j \rangle} J_{ij}\,{\mathbf S}_i \cdot
{\mathbf S}_j
\label{eq:1}
\end{equation}
The bonds are randomly taken from a quenched, binary probability
distribution,
\begin{equation}
P(J_{ij})= p\,\delta (J_{ij}-J_0)+ (1-p)\,\delta (J_{ij}+J_0)\ , 
\label{eq:2}
\end{equation}
so for $p=1/2$ one has the unmagnetized spin glass.
 
The Mattis model ascribes disorder to sites rather
than bonds ($J_{ij} \to J_0\,\zeta_i\,\zeta_j$), so that the Hamiltonian
reads:
\begin{equation}
{\cal H}_M= -J_0\sum_{\langle i,j \rangle} \zeta_i\,\zeta_j\,{\mathbf S}_i
\cdot {\mathbf S}_j\ ,
\label{eq:3}
\end{equation}
where $\zeta_i=+1$ ($-1$) with probability $p$ ($1-p$).
This way, the overall energy is minimized by making $S_i^z =\zeta_i\,S$,
which constitutes a (classical) ground state of the Hamiltonian
Eq.~(\ref{eq:3}), to be referred to as $|\,0\rangle$.
Thus, disorder is effectively removed from static properties, but not from
the dynamics, because of the handedness of Heisenberg spin commutation
relations. Indeed, considering low-energy excitations, the equations of
motion for the spins are, with $\hbar=1$: 
\begin{equation}
i\, dS_i^-/dt =\sum_j
J_0\,\zeta_i\,\zeta_j\,\left(S_i^-\,S_j^z -S_j^-\,S_i^z\right)\ ,
\label{eq:4} 
\end{equation}
where $S_i^\pm = S_i^x \pm i\,S_i^y$ etc and $j$ are nearest
neighbors of site $i$. So, putting $v_i \equiv \zeta_i\,S_i^-$,  one 
gets~\cite{sp88}, upon application
of Eq.~(\ref{eq:4}) to $|\,0\rangle$~:
\begin{equation}
i\,\zeta_i\,d u_i/dt = \sum_j J_0\,\left(u_i -u_j\right)\ .
\label{eq:5}
\end{equation}
where the $u_i$ are Mattis-transformed local (on-site) spin-wave
amplitudes.
For the eigenmodes with frequency $\omega$ (in units of the exchange
constant $J_0$), Eq.~(\ref{eq:5}) leads to
\begin{equation}
\omega\,\zeta_i\,u_i=  \sum_j \left(u_i -u_j\right)\ .
\label{eq:6}
\end{equation}
Goldstone modes are expected to occur, since disorder does not
destroy the symmetry of the system in spin space~\cite{gc03}. The
relationship of frequency to wave number, $k$, at low energies is
characterized by the dynamic exponent $z$:
\begin{equation}
\omega \propto k^z\ .
\label{eq:7}
\end{equation}
In $d=1$, where the scattering length coincides with the
localization length~\cite{ga04}, the definition of $k$ is
unique. Indeed, numerical calculations~\cite{bh89,ew92,ah93} of the $d=1$
density of states  and of the Lyapunov exponent point to the same value 
$z=3/2$, predicted  analytically~\cite{sp88}. For $d>1$ this degeneracy is
expected to be lifted. As we shall see below, different exponents come up,
depending on  whether localization or density-of-states properties are
being considered.  

\section
{Scaling}
\label{sec:3}
We briefly review the treatment of one-dimensional systems, given in
Ref.~\onlinecite{sp88}. In this case, Eq.~(\ref{eq:6}) becomes
\begin{equation}
(2-\zeta_i\,\omega)\,u_i = u_{i-1}+u_{i+1}\ .
\label{eq:1d}
\end{equation}
A transfer-matrix (TM) approach\cite{hori,ps81}, can be formulated,
giving~\cite{sp88,bh89,ew92}:
\begin{equation}
\begin{pmatrix}{u_{i+1}}\cr{u_i}\end{pmatrix}=
\begin{pmatrix}{2-\zeta_i\,\omega}&{-1}\cr{1}&{0}\end{pmatrix}
\begin{pmatrix}{u_i}\cr{u_{i-1}}\end{pmatrix} = T_i(\omega)\,
\begin{pmatrix}{u_i}\cr{u_{i-1}}\end{pmatrix}\ . 
\label{eq:1dtm}
\end{equation}
The allowed frequencies for a chain with $N$ spins 
and periodic boundary conditions, $u_{N+1} \equiv u_1$,
are determined by ${\rm
det}\,(\Lambda_N -1)=0$, where
\begin{equation}
\Lambda_N(\omega) =\prod_{i=1}^N T_i(\omega)\ ;
\label{eq:1dtm2}
\end{equation}
equivalently, the condition ${\rm Tr}\,\Lambda_N =2$ determines
the eigenfrequencies. Scaling the system by a linear dilation factor $b$,
the dynamics is preserved if the frequencies are transformed
($\omega \to \omega^\prime$), in such a way that 
\begin{equation}
{\rm Tr}\,\Lambda_N(\omega)={\rm Tr}\,\Lambda_{N/b}(\omega^\prime)\ .
\label{eq:1dtm3}
\end{equation}
Using properties of the matrices $T_i(\omega)$, one
finds~\cite{sp88} that the first-order term (in $\omega$) of ${\rm
Tr}\,\Lambda_N(\omega)$ has a coefficient equal to $N\sum_{i=1}^N \zeta_i$.
Therefore, correspondence of the $\{\zeta_i\}$ with an unbiased random-walk
makes the determining variable $\omega\,N^{3/2}$, so that the (length)
scaling of the frequencies is $\omega^\prime=\omega\,b^{3/2}$, and the
low-energy dispersion relation Eq.~(\ref{eq:7})
has an anomalous power (dynamic exponent) $z=3/2$.
In fact, careful consideration of higher-order terms~\cite{sp88} shows that
the combination $N^{3/2}\,\omega$ is present to all orders, thus scaling is
expected to hold even away from the $\omega \to 0$ region (though not the
single power-law form,  Eq.~(\ref{eq:7})).   

A suitable framework for extensions of this treatment to space
dimensionalities $d >1$ is found 
in quasi-- one dimensional geometries, i.e. $L^{d-1}
\times N$ systems with $N \gg 1$. In what follows, we shall always make use
of periodic boundary conditions across the $d-1$ transverse directions.

Considering $d=2$ for simplicity, a TM can be set up on a strip of width
$L$ sites, so an $L$- component vector ${\vec u}_i=(u_{1i}, \cdots ,
u_{Li})$ corresponds to each column $i$ along the strip, with the
recursion relation   
\begin{equation}
\begin{pmatrix}{{\vec u}_{i+1}}\cr{{\vec u}_i}\end{pmatrix}=
T_i^{2d}(\omega)\,
\begin{pmatrix}{{\vec u}_i}\cr{{\vec u}_{i-1}}\end{pmatrix}\ , 
\label{eq:2dtm}
\end{equation}
where
\begin{equation}
T_i^{2d}(\omega) =
\begin{pmatrix}{M_i}&{-I}\cr{I}&{0}\end{pmatrix}\,\quad
M_i = a -\omega b_i\ ,
\label{eq:2dtm2}
\end{equation}
$I$ being the $L \times L$ identity matrix, while $a$ and $b_i$ are given
by:
\begin{equation}
{a}=\begin{pmatrix}{4}&{-1}&{\ \, 0}&{ \cdots}&{
-1}\cr
{-1 }&{ 4}&{ -1}&{\cdots }&{\ 0}\cr
{\cdots }&{ \cdots }&{ }&{ }&{ -1}\cr
{-1 }&{\ 0 }&{ \cdots  }&{-1 }&{\ 4}\end{pmatrix}\,;\,b_i=
\begin{pmatrix}{\zeta_{1i}}&{\ \, 0}&{ \cdots}&{0}\cr
{0 }&{ \zeta_{2i}}&{ 0}&{\cdots }\cr
{\cdots }&{ \cdots }&{ }&{0 }\cr
{\ 0 }&{ \cdots  }&{0 }&{\ \zeta_{Li}}\end{pmatrix}
\label{eq:2dtm3}\,.
\end{equation}
Hence,
\begin{equation}
T_i^{2d}(\omega) =\begin{pmatrix}{a}&{-I}\cr{I}&{0}\end{pmatrix}-
\omega \begin{pmatrix}{b_i}&{0}\cr{0}&{0}\end{pmatrix} \equiv
A -\omega\,B_i\ .
\label{eq:2dtm4}
\end{equation}
Generalizations to higher $d$ are immediate, with the vector ${\vec u}_i$
now having $L^{d-1}$ components, and the matrices $I$, $a$ and $b_i$
being $L^{d-1} \times L^{d-1}$. 
The $(2L^{d-1} \times 2L^{d-1})$ matrix $T_i$ is symplectic, that
is, its eigenvalues occur in pairs $\{ \nu_i, \nu_i^{-1}\},\
i=1, \dots, L^{d-1}$. Note that matrix $A$ is symplectic as well.

For $d>1$, a feature which does not occur in the one-dimensional case is
that there are transverse momentum modes.
Returning to $d=2$ for illustration, these are indeed the eigenmodes of
matrix $a$ in
Eq.~(\ref{eq:2dtm3}), with corresponding energies $\varepsilon_p=
4-2\cos 2\pi p/L$, $p=0, 1, \dots, L-1$.

We briefly make contact with the analogous case of a homogeneous system of
length $N$, for which $b=I$, $[a,b]=0$, and
the eigenstates of $a$ are also eigenstates of the full hamiltonian, with
$\omega_{pq}=\varepsilon_p-2\cos 2\pi q/N$, $q=0, 1, \dots, N-1$.
This reminds us that, in $d>1$, the energy of a mode is not related only to
its longitudinal wavevector, as is the case in $d=1$. Upon introduction of
randomness, the commutation relation is destroyed (contrary to the
one-dimensional case where both $a$ and $b$ are numbers) and, consequently,
the interplay between frequency- and wavevector- aspects can only be
measured via the accumulated statistics of many local realizations of
disorder. Therefore, in $d>1$ one may expect the picture of a single
length controlling both (spatial) attenuation and (time) oscillation
damping~\cite{sp88}, which holds for $d=1$ spin glasses, to be replaced by
one where each of these properties is governed by a distinct quantity. 

We now return to spin glasses. 
From the eigenvectors of $a$, ``spinor'' generalizations can be built,
which are 
eigenvectors of $A$, with eigenvalues $(\nu_p,\nu_p^{-1})$ indexed by $p$;
one can show that $\nu_p+\nu_p^{-1}=\varepsilon_p$. While such spinors
are obviously not eigenvectors of $B$,  the contribution given by each
diagonal element of $T_i^{2d}(\omega)$, corresponding to fixed $p$, to the 
trace of $\Lambda_N^{2d}(\omega) \equiv \prod_{i=1}^N T_i^{2d}(\omega)$,
can be worked out to first order in $\omega$. Use is made of the fact
that, analogously to the $d=1$ case~\cite{sp88},
\begin{equation}
\prod_{\ell=1}^N T_\ell^{2d}(\omega) =
A^N-\omega\,\sum_{\ell=1}^N A^{\ell-1}\,B_{\ell}\,A^{N-\ell} + {\cal
O}(\omega^2)\ .
\label{eq:expt}
\end{equation}
The result is: 
\begin{eqnarray}
\nonumber
{\rm Tr}\,(p)\prod_{\ell=1}^N T_\ell^{2d}(\omega)  =\nu_p^N+\nu_p^{-N} -\\
-\omega\,\frac{(\nu_p^N-\nu_p^{-N})}{\nu_p-\nu_p^{-1}}\sum_{\ell=1}^N
\sum_{m=1}^L \frac{1}{L}\,\zeta_{m\ell} +{\cal O}(\omega^2)\ ,
\label{eq:trace}
\end{eqnarray}
where ${\rm Tr}\,(p)$ denotes the joint contribution of both eigenspinors
of $A$ indexed by $p$ (associated respectively to eigenvalues
$\nu_p$ and $\nu_p^{-1}$).

The ``critical'' (large scale) behavior is associated with
small $p$, in which case  $\nu_p,\ \nu_p^{-1} \to 1$, and
Eq.~(\ref{eq:trace}) turns into:
\begin{equation}
{\rm Tr}\,(p)\prod_{\ell=1}^N T_\ell^{2d}(\omega) \to 2
-\omega\,N\sum_{\ell=1}^N \frac{1}{L}\,\sum_{m=1}^L\zeta_{m\ell} 
+{\cal O}(\omega^2)\ .
\label{eq:lowp}
\end{equation}

One can readily see that, for generic $d>1$, this translates into:
\begin{eqnarray}
\nonumber
{\rm Tr}\,(p)\prod_{\ell=1}^N T_\ell(\omega) \to 2 -\\
-\omega\,N\sum_{\ell=1}^N\left(\prod_{i=1}^{d-1} 
\frac{1}{L_i}\,\sum_{m_i=1}^{L_i}\,\zeta_{m_1\cdots m_{d-1}\ell}\right)
+{\cal O}(\omega^2)\ .
\label{eq:lowpd}
\end{eqnarray}
In the second term of Eq.~(\ref{eq:lowpd}), one has a sum of $N\times
L_1\times  \cdots \times L_{d-1}$ binary random variables, so this is
gaussian distributed with rms value:
\begin{equation}
\frac{\omega\,N}{\prod_{i=1}^{d-1}L_i}\,\left(N
\prod_{i=1}^{d-1}L_i\right)^{1/2} = 
\frac{\omega\,N^{3/2}}{\left(\prod_{i=1}^{d-1}L_i\right)^{1/2}}\ .   
\label{eq:sqrt}
\end{equation}
Upon scaling of linear dimensions by a factor $b$, under which frequency
scales as $\omega \to b^z\,\omega$, and requiring invariance of the
term given in Eq.~(\ref{eq:sqrt}) (see Eq.~(\ref{eq:1dtm3})), one gets:
\begin{equation}
z=2-\frac{d}{2} =\begin{cases}\frac{3}{2} & d=1\cr 1 & d=2\cr
\frac{1}{2} & d=3\cr \end{cases}\ .
\label{eq:zlowp}
\end{equation}
Consideration of the terms in Eq.~(\ref{eq:lowp}) of higher than first
order in $\omega$ shows that, unlike the $d=1$ case, the trace of the 
full TM is not just a function of the variable given in
Eq.~(\ref{eq:sqrt}), because complicated sums occur, involving both
longitudinal and  transverse wave vectors. 

This is in line with the 
reasoning presented above, to the effect that the simultaneous
presence of both longitudinal and transverse degrees of freedom invalidates 
the single-length picture, predicted analytically~\cite{sp88}
and numerically confirmed~\cite{ah93,bh89,ew92} for $d=1$.

While it is plausible to expect that, for
some low-energy regime in $d>1$ the scaling result, Eq.~(\ref{eq:zlowp})
might hold true, direct
verification is called for. 

\section{Numerical analysis}
\label{sec:4}
\subsection{Lyapunov exponents}
\label{subsec:lyap}
The procedure for calculating Lyapunov exponents on strips or bars
is the same as that used for Anderson localization problems~\cite{ps81}.
Indeed, in both cases the TM is symplectic, and one can use Oseledec's
theorem and dynamic filtration to extract the smallest Lyapunov exponent,
whose inverse is the largest localization length. For Heisenberg
spin-glass chains, this has been done~\cite{ew92,bh89}, numerically
confirming
the result $z=3/2$ obtained analytically in Ref.~\onlinecite{sp88}. 
 
We have investigated strips of widths $L=4,6,\cdots, 14$ in $d=2$, in
which for each energy $\omega$ we took $N=10^6$ iterations of the TM, and
bars with $L \times L$ cross-section, $L=4,6,8,10$ in $d=3$.
In $d=3$ we used $N=10^6$ for $L=4, 6$, $5 \times 10^5$ for $L=8$, 
and $1 \times 10^5$ for $L=10$.

In contrast with $d=1$, here one must take into account finite-size
effects, introduced via the transverse dimension $L$, thus
calculated localization lengths are denoted by $\lambda_L$. Using
standard finite-size scaling theory~\cite{fs2}, it is expected that
the behavior of scaled localization lengths $\lambda_L/L$, when
plotted against $\omega\,L^z$, will allow one to infer the bulk
($L \to \infty$) properties of the system.
\begin{figure}
{\centering \resizebox*{3.3in}{!}{\includegraphics*{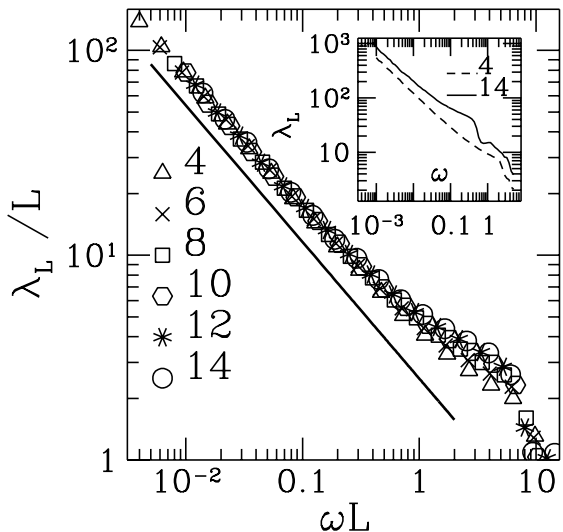}}}
\caption{Scaling plot for localization lengths on strips of a
$d=2$ system, against $\omega\,L^z$, with $z=1$ as predicted
in Eq.~(\protect{\ref{eq:zlowp}}). Strip widths $L$ as given by symbols. 
Line corresponds to $y \propto x^{-2/3}$ and is a guide to the eye,
showing how the effective $d=1$ regime sets in for very low energies. 
Insert: unscaled data for $L=4$ and $14$.
} 
\label{fig:lce2d}
\end{figure}

In Fig.~\ref{fig:lce2d} we see that in $d=2$ good data collapse, extending
as far as $x \equiv \omega\,L^z \simeq 0.3$, is achieved 
when $z=1$, as predicted in Eq.~(\ref{eq:zlowp}). At the
low-energy end, $x \lesssim 0.03$, the quasi-one dimensional
character of the strips begins to dominate, and the scaling curve crosses
over to the effective $d=1$ regime characterized by $\omega \sim k^{3/2}$.

\begin{figure}
{\centering \resizebox*{3.3in}{!}{\includegraphics*{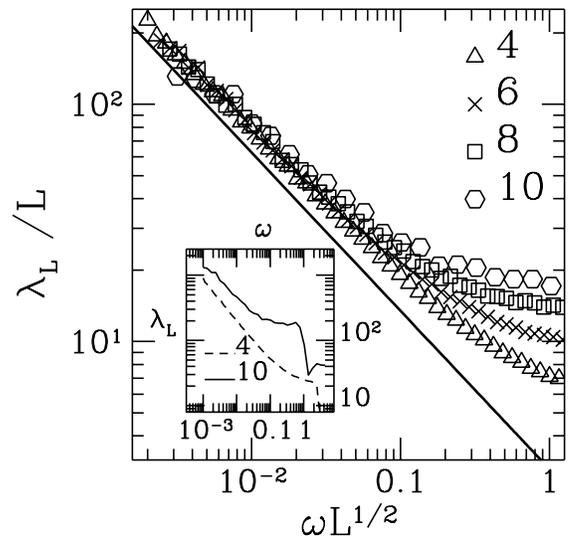}}}
\caption{Scaling plot for localization lengths on bars of a
$d=3$ system, against $\omega\,L^z$, with $z=1/2$ as predicted
in Eq.~(\protect{\ref{eq:zlowp}}). Bar cross-sections are
$L\times L$, with  $L$ as given by symbols. 
Line corresponds to $y \propto x^{-2/3}$ and is a guide to the eye,
showing how the effective $d=1$ regime sets in for very low energies. 
Insert: unscaled data for $L=4$ and $10$.
} 
\label{fig:lce3d}
\end{figure}
In Fig.~\ref{fig:lce3d} the scaling plot for $d=3$, with $z=1/2$ as 
predicted in Eq.~(\ref{eq:zlowp}), is exhibited. The quality of
data collapse is remarkably inferior to that of $d=2$ data. 
An examination of the behavior of $\lambda_L/L$ against $\omega$
shows that curves corresponding to pairs $L,L-2$ have well-defined
crossings at low energies $\omega \lesssim 0.05$. The usual interpretation
of these, in
the finite-size scaling context, would point to a
localization-delocalization transition~\cite{ps81,fs2}. However, we
have found that the locations of crossings appear to approach 
$\omega =0$ with increasing $L$. This would be consistent with
the idea that all magnons are delocalized in $d=3$, which is supported,
e.g., by the field-theoretical results of Ref.~\onlinecite{ga04}. We
postpone a  discussion of this point (and similar ones associated
to the behavior found above for $d=2$), to Section~\ref{sec:conc}.

\subsection{Densities of states}
\label{subsec:dos}

The calculation of densities of states per unit energy interval (DOS),
${\cal D}(\omega)$, and their integrated counterparts (IDOS),
$N(\omega)=\int_{-\infty}^\omega
{\cal D}(\omega^\prime)\,d\omega^\prime$, makes use of eigenvalue-counting
theorems~\cite{dean,dm60,tho72}. Our implementation resorts
to Gaussian elimination algorithms on quasi one-dimensional
geometries ($L^{d-1}
\times N$, with $ N \gg L$), and closely follows the
steps described in Refs.~\onlinecite{sb83,sne86} where the systems under
investigation were, respectively, phonons in disordered
solids, and tight-binding electrons (Anderson localization). The key
feature shared between these problems and the one studied here is
the fact that, for an $L^{d-1} \times N$ system with periodic boundary
conditions across, the hamiltonian has a $(2\times L^{d-1}+1)$- diagonal
form,  i.e., it can only have non-zero elements in the $L^{d-1}$ lines
above, and $L^{d-1}$ lines below, the diagonal.

We consider the characteristic matrix, which in the present case is
$C=\zeta\omega\,I-{\cal H}$, where $\zeta\omega\,I$ is a diagonal matrix
with $[\zeta\omega\,I]_{jj} =\zeta_j\,\omega$ ($j =$ site index),
for an $L^{d-1} \times N$ system. Evaluation of its diagonal elements via
Gaussian elimination  enables one to obtain the IDOS for any
energy~\cite{sb83,sne86}, thus the DOS may be calculated by numerical
differentiation.

\begin{figure}
{\centering \resizebox*{3.3in}{!}{\includegraphics*{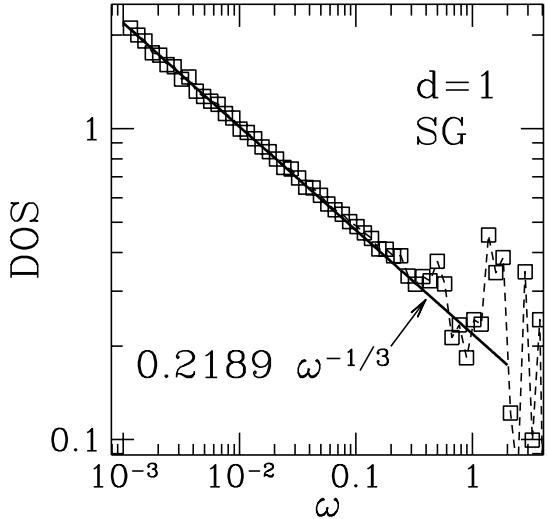}}}
\caption{Double-logarithmic plot of density of states ${\cal D}(\omega)$
for spin-glass chain,
calculated by Gaussian elimination. Chain length $N=10^7$ sites. Thick
line is the exact Derrida-Gardner result~\protect{\cite{dg84}} (with
coefficient doubled, on account of different normalization, see text).
} 
\label{fig:dos1d}
\end{figure}

For $d=1$ the eigenvalue counting (used., e.g., in Ref.~\onlinecite{ew92}),
may, alternatively, proceed via enumeration of nodes of the amplitude
ratios which enter the
evaluation of the (single) Lyapunov exponent~\cite{bh89,ah93}. In order
to test our recursion and elimination algorithms, we applied them to this
case  and compared the outcome with that from node-enumeration . 
Results are identical to within numerical accuracy, and
the set produced by Gaussian elimination is depicted in
Fig.~\ref{fig:dos1d}.
By sampling energies separated by logarithmically uniform intervals, we
achieved a detailed view of the $\omega \to 0$ region, which is 
difficult to isolate in the corresponding DOS results of 
Refs.~\onlinecite{bh89,ew92,ah93} (where linear binning was used). One sees
that the relationship ${\cal
D}(\omega) \propto \omega^{-1/3}$ is valid for more than two orders of
magnitude in energy, up
to $\omega \simeq 0.3$. For guidance, we have also included the exact
Derrida-Gardner result~\cite{dg84}. Since we have considered only positive
energy states in our calculation, the appropriate
proportionality coefficient is twice
that given in Ref.~\onlinecite{dg84}.

For higher-dimensional cases, is is worth mentioning that the algorithms
used here are much less computationally intensive than their
Lyapunov-exponent counterparts. For an $L^{d-1}\times N$ system, the
computational time rises as $L^{3(d-1)}\times N$ for the
former~\cite{sne86}, and approximately as $L^{5(d-1)}\times N$ for the
latter. This is mainly because of the frequent mutual orthogonalization of
$2\times L^{d-1}$ iterated vectors, which is necessary in order
to avoid cross-contamination between eigenvectors associated to
different Lyapunov exponents. Therefore, for DOS and IDOS  
it is usually possible (except for very low energies in $d=3$, see below)
to work with systems whose transverse dimensions $L$ are large enough 
that finite-size effects are of little import. It remains only
to  make sure that the sample length  $N$ is long enough, in order to
achieve adequate sampling of quenched disorder configurations.

We examined the effect of finite transverse dimensions, by evaluating
pure-system quantities and comparing our results to the exact ones.
Though, having zero net magnetization, the spin glasses studied here are
closer to antiferromagnets (AF) than to
homogeneous ferromagnets (FM), the DOS and IDOS of magnons in the latter
exhibit some distinctive features, whose numerical reproduction is a
non-trivial test of the adequacy and accuracy of our methods. 
For FM in $d=2$, already with $L=25$, $N=2500$ the IDOS is at most $3\%$
off the exact value. This largest discrepancy happens close to $\omega=4$
where the analytical
IDOS exhibits an inflection point, on account of the DOS's logarithmic Van
Hove singularity at the
band center. Increasing $L$ or $N$ does not significantly reduce
the deviation close to $\omega=4$; however, it does improve agreement
elsewhere on the energy axis.
The calculated DOS is rather sensitive to discrete-lattice
effects; nevertheless, the consequent oscillations are again much
diminished by increasing  $L$, $N$. For $d=3$ FM, the relatively
featureless IDOS is easier to reproduce. With $L=16$, $N=25600$, deviations
are down to, at most, $1.5\%$ (though the DOS still displays somewhat large
oscillations, especially around the ``knees'' at $\omega=4$ and $8$). 
\begin{figure}
{\centering \resizebox*{3.3in}{!}{\includegraphics*{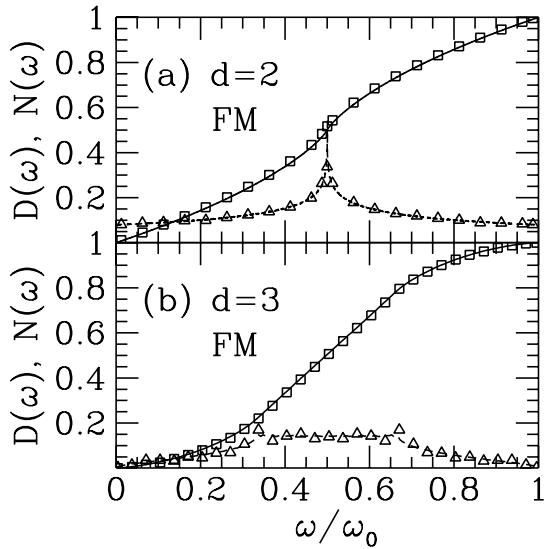}}}
\caption{DOS (${\cal D}(\omega)$) and IDOS ($N(\omega)$) for pure FM
systems against normalized energy $\omega/\omega_0$: analytical (lines)
and calculated by Gaussian elimination
(points). Dashed lines and triangles: ${\cal D}(\omega)$; full lines and
squares: $N(\omega)$. (a): $d=2$, $L=100$, $N=50000$. (b): $d=3$, $L=16$,
$N=25600$. The pure-system (FM) bandwidth is $\omega_0^F = 8$ $(d=2)$, and
$12$ $(d=3)$.
} 
\label{fig:idosp}
\end{figure}
Fig.~\ref{fig:idosp} shows representative results, which are useful as
guidelines for the investigation of disordered systems in $d=2$ and $3$
via Gaussian elimination. 

\begin{figure}
{\centering \resizebox*{3.3in}{!}{\includegraphics*{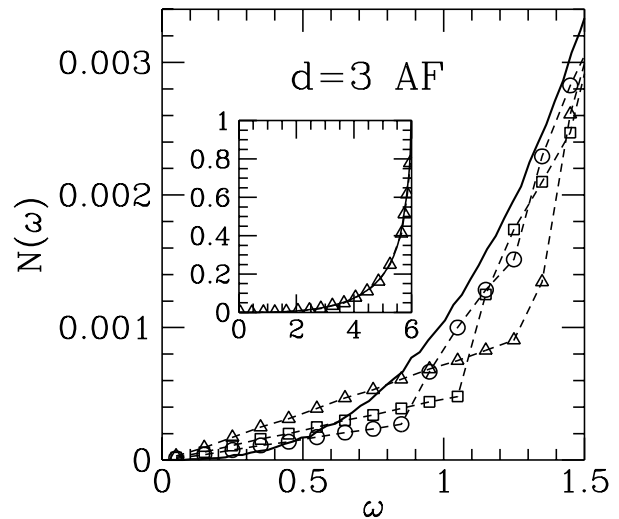}}}
\caption{Low-energy IDOS, $N(\omega)$, for pure AF in $d=3$
against  energy $\omega$: analytical (full line)
and calculated by Gaussian elimination
on $L^2 \times N$ systems, $N=500\,L^2$ (points, connected by dashed
lines).  Triangles: $L=16$; squares: $L=20$; circles: $L=24$. Inset:
full-band IDOS (same axes as main Figure). Analytical (full line),
and Gaussian elimination with $L=16$ (triangles).
} 
\label{fig:dosaf3d}
\end{figure}

Turning to pure AF systems, for which the respective bandwidths are
$\omega_0^{AF} = 4$ $(d=2)$, and $6$ $(d=3)$, again
relatively small transverse dimensions $L$ provide results which closely
follow the analytic values, except at very low $\omega$. In this limit, the
fact that the finite $L$ quantizes the transverse momentum  leads to 
effective one-dimensional behavior (${\cal D}(\omega) \sim
\omega^0$, $N(\omega) \sim \omega^1$) for $\omega$ less than a crossover 
frequency $\omega_m \equiv A_{\,\rm AF}(d)/L^z$, $z=1$.
With the units used in this work, we found $A_{\,\rm AF}(2) \simeq 12$,
$A_{\,\rm AF}(3) \simeq 20$. The effect is more pronounced here than for
FM, where $z=2$
and, consequently, the onset of this sort of behavior occurs at much lower
energies. Fig.~\ref{fig:dosaf3d} highlights the worst case of $d=3$.
For completeness, the inset of Fig.~\ref{fig:dosaf3d} shows that, even
for $L=16$ where these low-energy discrepancies are rather severe,
agreement with analytical forms is quite satisfactory elsewhere.    

\begin{figure}
{\centering \resizebox*{3.3in}{!}{\includegraphics*{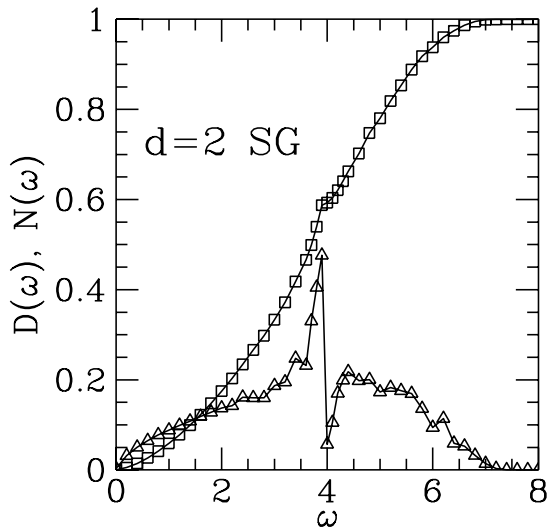}}}
\caption{DOS (${\cal D}(\omega)$) and IDOS ($N(\omega)$) for Mattis spin 
glass in $d=2$, against energy $\omega$, calculated by Gaussian
elimination. 
Triangles: ${\cal D}(\omega)$; squares: $N(\omega)$.  $L=250$, $N=2.5
\times 10^6$.
} 
\label{fig:dos2d}
\end{figure}

We now return to disordered systems.
In Fig.~\ref{fig:dos2d}, results for the Mattis spin glass in $d=2$
are presented. We used $L=250$, $N=2.5 \times 10^6$. The
number of sites entering the calculation was more than one order of
magnitude larger than in that for a pure FM, whose result is exhibited
in Fig.~\ref{fig:idosp}~(a). From examination of shorter runs for the
disordered case, it appears that the features displayed in
Fig.~\ref{fig:dos2d} are rather stable and well-converged. 
For this value of $L$, the crossover to one-dimensional behavior,
referred to above, is confined to $\omega \lesssim 0.05$, leaving a
broad window at low energies for which genuine two-dimensional
behavior can be observed. The main
distinctions of the IDOS from its pure-system (FM and AF) 
counterparts are:
(i) close to $\omega =4$, the upper limit of the AF band, the FM IDOS's
inflection point is replaced by a seeming ``knee'', with a short flat
section; and (ii)
saturation is reached below the FM band edge $\omega_0^F=8$, but
above the AF edge $\omega_0^{AF}=4$;
by $\omega=6.7$ the IDOS is already within less than $1\%$ of unity.
Similar effects can be seen in early numerical work~\cite{ch79}, though
in that Reference saturation appears to be reached only above the FM band
edge, at $\omega \simeq 9.0$.  

It has been predicted~\cite{ds79,ga04} that, since $d=2$ is the  critical
dimensionality in this case~\cite{gc03}, the two-dimensional spin glass
will behave as a pure (AF) system (namely, ${\cal D}(\omega) \sim
\omega^1$, $N(\omega) \sim \omega^2$), with logarithmic corrections.
At low frequencies, the real part of the dispersion relation is expected to
follow the expression~\cite{ds79,ga04}:
\begin{equation}
{\rm Re}\ \omega \propto
\frac{p}{\sqrt{\log\left(\frac{\Lambda}{p}\right)}}\ ,
\label{eq:g2d}
\end{equation}
where $\Lambda$ is a momentum cutoff, reciprocal to the minimum wavelength
of magnons. From Eq.~(\ref{eq:g2d}), one can work out the predicted 
behavior of the IDOS at low energies. This turns out to be:
\begin{equation}
N(\omega) \propto \omega^2\,\ln \left(\frac{\Omega}{\omega}\right)\ ,
\label{eq:2dcorr}
\end{equation}
where $\Omega$ is a cutoff frequency, corresponding to the momentum cutoff
$\Lambda$.

\begin{figure}
{\centering \resizebox*{3.3in}{!}{\includegraphics*{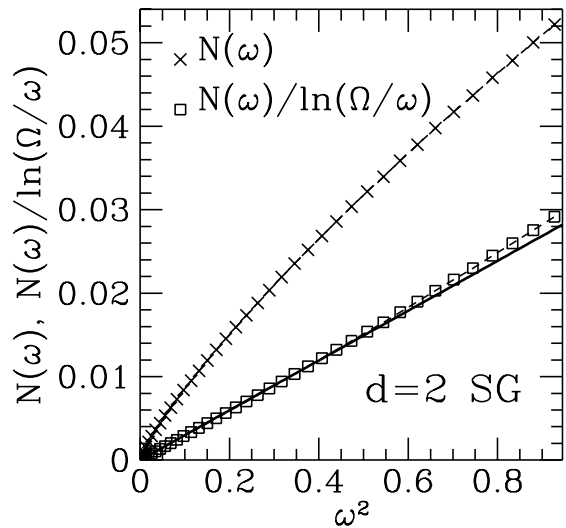}}}
\caption{IDOS  ($N(\omega)$) for Mattis spin  glass in $d=2$, for low
energies, against $\omega^2$, calculated by Gaussian elimination.  
$L=250$, $N=2.5 \times 10^6$. 
Crosses: $N(\omega)$; squares: $N(\omega)/\ln(\Omega/\omega)$,
with $\Omega=5.8$ [see Eq.~(\protect{\ref{eq:2dcorr}}), and text].
Full line is a linear least-squares fit to data for $0.05 \leq \omega \leq 
0.5$.
} 
\label{fig:dos2dle}
\end{figure}

We have tested the prediction, Eq.~(\ref{eq:2dcorr}), against our data,
with the results shown in Fig.~\ref{fig:dos2dle}. A fit of the raw data
(crosses in Fig.~\ref{fig:dos2dle}) to pure power-law behavior gives
$N(\omega) \sim \omega^x$, with the effective exponent $x \simeq 1.62$. 
On the other hand, plotting $N(\omega)/\ln(\Omega/\omega)$ against 
$\omega^2$ (squares in Fig.~\ref{fig:dos2dle})  removes just about all the
curvature, provided that a suitable value of $\Omega$ is used. A
linear least-squares fit of data for $0.05 \leq \omega \leq 0.5$
(shown as a full line in Fig.~\ref{fig:dos2dle})
gives $\Omega =5.8(1)$, broadly consistent with the effective bandwidth
$\gtrsim 6.7$ found above . Keeping $\Omega =5.8$, and fitting
$N(\omega)/\ln(\Omega/\omega)$ to a power law dependence  over the full
interval $0.05 \leq \omega \leq 1.0$, would give an effective power $x
\simeq 1.04$.

We undertook similar calculations for the Mattis spin glass in $d=3$.
Since one is above the  critical dimensionality in this
case~\cite{ds79,gc03},
the three-dimensional spin glass is expected to behave as a pure (AF)
system, at least at low energies and long wavelengths (namely, ${\cal
D}(\omega) \sim \omega^2$, $N(\omega) \sim \omega^3$).

\begin{figure}
{\centering \resizebox*{3.3in}{!}{\includegraphics*{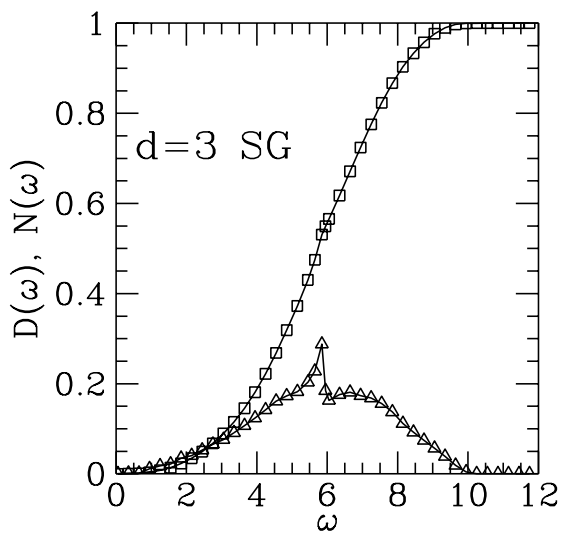}}}
\caption{DOS (${\cal D}(\omega)$) and IDOS ($N(\omega)$) for Mattis spin 
glass in $d=3$, against energy $\omega$, calculated by Gaussian
elimination. 
Triangles: ${\cal D}(\omega)$; squares: $N(\omega)$.  $L=16$, $N=2.56
\times 10^6$.
} 
\label{fig:dos3d}
\end{figure}

Similarly to the pure $d=3$ AF, for the ranges of $L$ within relatively
easy reach of our  calculations, the low-frequency spectrum exhibits a
crossover towards one-dimensional behavior. With the terminology introduced
above, this happens for $\omega \lesssim \omega_m$, $\omega_m
=A_{\,\rm SG}(d)/L$; by examining the sequence $L=16,\,24,\,30,\,36$, we
estimate
$A_{\,\rm SG}(3) \simeq 11$, just over half the corresponding value for
pure AF.  Thus, such effects are once more confined to low energies. We
have found that, for $\omega \gtrsim 1.2$, the $L=16$ curve 
is within less than $3\%$ of those corresponding to larger $L$,
which are grouped together even more tightly.  
Fig.~\ref{fig:dos3d} presents an overall picture of
results, for $L=16$, $N=2.56 \times 10^6$.
Again, early saturation occurs. The IDOS is within $0.1\%$ of unity by
$\omega =9.4$, just over three-quarters of the FM band width
$\omega_0=12$. A kink, similar to the one occurring in $d=2$ but less
intense, arises close to the center of the FM band (and top of the AF one),
$\omega =6$. Both features show up in Ref.~\onlinecite{clh77},
though with saturation occurring at a slightly higher energy (but still
within the FM band).

\begin{figure}
{\centering \resizebox*{3.3in}{!}{\includegraphics*{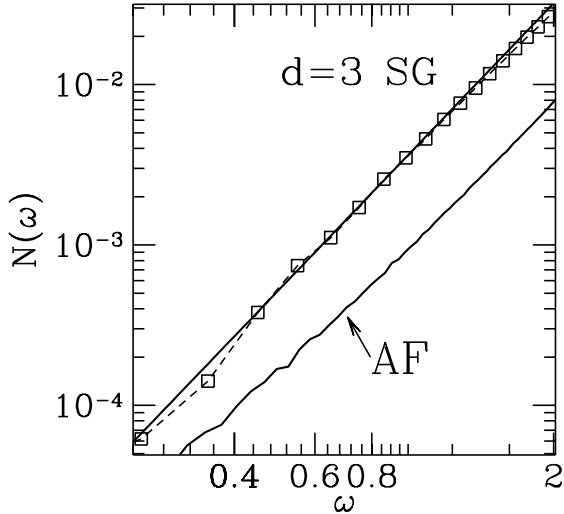}}}
\caption{Double-logarithmic plot of IDOS
($N(\omega)$) for Mattis spin  glass in $d=3$, for low energies, against
energy $\omega$, calculated by Gaussian elimination.  
Squares: calculated points, $L=36$, $N=6.48\times 10^5$. The straight
line is a power-law fit with slope $2.97$ (from least-squares fit of data
for $0.45 \leq \omega \leq 1.0$). Also shown is IDOS for pure AF,
calculated for a cube with $100^3$ sites.
} 
\label{fig:dos3dle}
\end{figure}

The low-energy behavior is shown in Fig.~\ref{fig:dos3dle}. For
$L=36$ we have found
that least-squares fits of our calculated data (excluding the very
low-energy intervals where one-dimensional behavior takes over) give
$z=2.97(5)$, if we keep to $\omega \leq 1.0$; including higher energies
(e.g. $\omega \lesssim 2.5-3.0$) results in a slight decrease of effective
exponents, down to $z \simeq 2.75$. On the other hand, fits of the
numerically-evaluated analytic IDOS for a cube 
with $100^3$ sites (shown in Fig.~\ref{fig:dos3dle}), when restricted to
$\omega \leq 1.5$, give an effective $z=2.82(1)$; it is only when the upper
limit is raised to $\omega=3.0$ that one reaches $z=2.99(1)$. This is
because, in the low-energy limit, discrete-lattice effects still persist,
which induce slight deviations of effective behavior away from the exact
value $z=3$. In summary, it is only in the very low-energy limit
$\omega \leq 1.0$ that the $d=3$ SG $N(\omega$) indeed exhibits the
$\omega^3$ dependence characteristic of the pure AF.

Therefore, we conclude that our low-energy data are consistent with the 
indications of Refs.~\onlinecite{gc03,ga04}, that magnons in  the $d=3$
Mattis SG display the same low-energy behavior as in a pure AF.
However, the respective amplitudes differ, as is apparent
by the roughly constant distance between SG and AF data in
Fig.~\ref{fig:dos3dle}. Writing $N_{\rm X}(\omega) =a_{\rm X}\,\omega^z$
(X=SG, AF),
we get from our fits: $a_{\,\rm SG}/a_{\,\rm AF}=4.1(1)$.

A calculation of the amplitudes, along the lines of Ref.~\onlinecite{ds79},
yields $a_{\,\rm SG}/a_{\,\rm AF}=(2I)^{3/2}=5.281\dots$, where
$I=1.516386\dots$ is Watson's integral~\cite{matbk}.
We believe the order-of-magnitude agreement found between our numerical
estimate and this result is satisfactory, given that disorder is
treated only approximately in the latter approach. 	

\section{Discussion and Conclusions} 
\label{sec:conc}
The preceding results are consistent with our statement, made in 
Sec~\ref{sec:2}, that the single-length picture
which prevails in $d=1$ cannot be ported to higher space dimensionalities.
In order to make contact with the one-dimensional case, we will refer to  
the indices emerging from the analytical scaling of
Sec.~\ref{sec:3}, and from the Lyapunov exponent calculations of 
Sec.~\ref{subsec:lyap} as $z_L$, while those originating from
the results of Sec.~\ref{subsec:dos} (plus the relationship $N(\omega)
\sim \omega^{d/z}$) will be denoted by $z_\omega$.

The analytical scaling predictions $z_L=1$ ($d=2$), $z_L=1/2$ ($d=3$),
are confirmed by our Lyapunov exponent calculations,
though the width of the energy intervals for which scaling holds is
larger for the former ($5 \times 10^{-3} \lesssim \omega\,L \lesssim 0.3$)
than for the latter  ($5 \times 10^{-3} \lesssim \omega\,L^{1/2} \lesssim
5 \times 10^{-2}$). 

In $d=2$, the curves of $\lambda_L/L$ against $\omega$
are essentially parallel for $\omega \lesssim 0.1$, down to the lowest
energies investigated; for fixed $\omega$, $\lambda_L/L$ decreases with
increasing $L$. This indicates the absence of a
delocalization transition, i.e. all modes are localized in 
$d=2$, in agreement with Refs.~\onlinecite{ga04,gc03}. On the other hand,
our result $z_L=1$ implies that the localization
length diverges at low energies as $\ell_{\rm loc} \sim \omega^{-1}$.
This is in contrast with the field-theoretical prediction of
Ref.~\onlinecite{ga04}, according to which $\ell_{\rm loc} \sim 
\omega^{-1/16 \pi}$.

For $d=3$, as mentioned above, the curves of $\lambda_L/L$ against
$\omega$ cross each other at low energies. For the $(L,L-2)=(6,4)$
pair, the crossing occurs at $\omega \simeq 0.04$, while for
$(10,8)$ it moves to lower energy $\omega \simeq 0.015$. We interpret
this as a residual finite-size effect, which will properly vanish with
increasing $L$, and see no reason why the established
idea~\cite{ga04,gc03} that all excitations are delocalized in $d=3$
should be challenged on the basis of such result.

A connection of our predictions for $z_L$ with the literature can be
made as follows. The analysis of Refs.~\onlinecite{ds79,ga04} was
carried out by assuming a well-defined (real) wavevector, thus implying the
complex dispersion relation:
\begin{equation} 
\omega (k) =\omega_R(k)+i\,\Gamma(k)\ .
\label{eq:disprel}
\end{equation} 
On the other hand, our TM formulation gives a specified (spatial) amplitude
decay ratio $\lambda^{-1}$ for a fixed (real) frequency, which then
envisages a complex wavevector,
\begin{equation}
k=k_R+i\,k_I\ ,\quad\lambda \sim k_I^{-1}\ . 
\label{eq:krki}
\end{equation}
One can then plug Eq.~(\ref{eq:krki}) back into Eq.~(\ref{eq:disprel}),
taking into account the specific dependencies of $\omega_R$ and $\Gamma$
on $k$, and force $\omega$ to be real in the latter.

For $d=3$, one expects~\cite{ds79,ga04} $\omega_R(k) \sim k$, $\Gamma
(k) \sim k^2$, consistent with small line broadening at low $k$ (i.e.
propagating modes). From this, one then gets:
\begin{equation}
\lambda^{-1} \sim \omega^2\quad (d=3)\ ,
\label{eq:z3d}
\end{equation}
so that the scaling variable is indeed $\omega L^{1/2}$.

For $d=2$, a similar argument can be made (now on somewhat flimsier
grounds, because all modes are expected to be localized, so the real and
imaginary parts of the dispersion relation may be of the same order
of magnitude). Ignoring logarithmic corrections, the results of
Refs.~\onlinecite{ds79,ga04} are: $\omega_R(k) \sim k$, $\Gamma
(k) \sim k$, from which we get:
\begin{equation}
\lambda^{-1} \sim \omega\quad (d=2)\ ,
\label{eq:z2d}
\end{equation}
again consistent with the $d=2$ scaling  variable being $\omega L$.    

The outcome of our density-of-states calculations for $d=2$ can be very
closely  fitted, for low energies $0.05 \leq \omega \leq 0.5$, to the
logarithmically-corrected form predicted in Ref.~\onlinecite{ga04} (see
Eqs.~(\ref{eq:g2d}),~(\ref{eq:2dcorr}), and Fig.~\ref{fig:dos2dle}).
Furthermore, one gets $z_\omega=1$ plus enhancing logarithmic corrections
(recall the effective exponent $\simeq 1.62$ from Fig.~\ref{fig:dos2dle}),
which is in line with the vanishing of group velocity 
(mode softening)~\cite{sp88} as $\omega \to 0$. 
  
Finally, our $d=3$ density-of-states results
are again consistent with the pure AF behavior
predicted~\cite{ds79,ga04,gc03}
to hold above $d_c=2$. Thus we have $z_\omega=1$ in this case.
However, the amplitudes of the low-energy power-law behavior differ, and we
have found $a_{\,\rm SG}/a_{\,\rm AF}=4.1(1)$.

\begin{acknowledgments}
We thank D. Sherrington, J. T. Chalker and Roger Elliott for interesting
discussions.
S.L.A.d.Q. thanks the Rudolf Peierls Centre for Theoretical Physics,
Oxford, where most of this work was carried out, for the hospitality,
and CNPq and Instituto do Mil\^enio de Nanoci\^encias--CNPq for 
funding his visit. The research of S.L.A.d.Q. was partially supported by
the Brazilian agencies CNPq (Grant No. 30.0003/2003-0), FAPERJ (Grant
No. E26--152.195/2002), FUJB-UFRJ, and Instituto do Mil\^enio de
Nanoci\^encias--CNPq.
R.B.S. acknowledges partial support from EPSRC Oxford Condensed Matter
Theory Programme Grant GR/R83712/01.
\end{acknowledgments}


\begin{thebibliography}{99}
\bibitem{dcm76}
D.C. Mattis, Phys. Lett. A {\bf 56}, 421 (1976).
\bibitem{ds77}
D. Sherrington, J. Phys. C {\bf 10}, L7 (1977).
\bibitem{ds79}
D. Sherrington, J. Phys. C {\bf 12}, 5171 (1979).
\bibitem{clh77}
W.Y. Ching, K.M. Leung, and D.L. Huber, \prl {\bf 39}, 729 (1977).
\bibitem{ch79}
W.Y. Ching and D.L. Huber, \prb {\bf 20}, 4721 (1979).
\bibitem{sp88}
R.B. Stinchcombe and I.R. Pimentel, \prb {\bf 38}, 4980 (1988).  
\bibitem{ga04}
V. Gurarie and A. Altland, J. Phys. A {\bf 37}, 9357 (2004).
\bibitem{hs77}
B.I. Halperin and W.M. Saslow, \prb {\bf 16}, 2154 (1977).  
\bibitem{ahc92}
I. Avgin, D.L. Huber, and W.Y. Ching, \prb {\bf 46}, 223 (1992).
\bibitem{ah93}
I. Avgin and D.L. Huber, \prb {\bf 48}, 13$\,$625 (1993).
\bibitem{ahc93}
I. Avgin, D.L. Huber, and W.Y. Ching, \prb {\bf 48}, 16$\,$109 (1993).
\bibitem{ew92}
S.N. Evangelou and A.Z. Wang, J. Phys.: Condens. Matter {\bf 4}, L617
(1992).
\bibitem{bh89}
A. Boukahil and D.L. Huber, \prb {\bf 40}, 4638 (1989).
\bibitem{hori}
J. Hori, {\it Spectral Properties of Disordered Chains and Lattices}
(Pergamon Press, Oxford, 1968). 
\bibitem{ps81}
J.-L. Pichard and G. Sarma, J. Phys. C {\bf 14}, L127 (1981); {\it ibid.}, 
{\bf 14}, L617 (1981).
\bibitem{sb83}
S. Baer, J. Phys. C {\bf 16}, 6939 (1983).
\bibitem{sne86}
S.N. Evangelou, J. Phys. C {\bf 19}, 4291 (1986).
\bibitem{gc03}
V. Gurarie and J.T. Chalker, \prb {\bf68}, 134207 (2003).
\bibitem{fs2}
M. P. Nightingale, in {\it Finite Size Scaling and Numerical 
Simulations of Statistical Systems}, edited by V. Privman (World
Scientific, Singapore, 1990). 
\bibitem{dean} 
P. Dean, Proc. Phys. Soc. London {\bf 73}, 413 (1959).
\bibitem{dm60}
P. Dean and J.L. Martin, Proc. Roy. Soc. A {\bf 259}, 409 (1960).
\bibitem{tho72}
D.J. Thouless,  J. Phys. C {\bf 5}, 77 (1972).
\bibitem{dg84}
B. Derrida and E. Gardner, J. Phys. (Paris) {\bf 45}, 1283 (1984).
\bibitem{matbk}
D.C. Mattis, {\it The Theory of Magnetism}
(Harper \& Row, New York, 1965), pg. 146. 
\end{thebibliography}
\end{document}